\documentclass[twocolumn,showpacs,superscriptaddress,amssymb,aps,prl]{revtex4-1}
\usepackage{amsmath,amsfonts,amssymb,amsthm,graphics,graphicx,epsfig,bbm}
\usepackage[colorlinks=true,citecolor=blue,linkcolor=blue,urlcolor=blue]{hyperref}
\usepackage[usenames]{color}
\usepackage{graphicx}
\usepackage{subfigure}
\usepackage{amsmath}
\usepackage{epsfig}
\usepackage{dcolumn}
\usepackage{bm}
\usepackage{color}
\usepackage{epstopdf}
\usepackage{amssymb}
\usepackage{amstext}
\usepackage{latexsym}
\usepackage{hyperref}
\usepackage{psfrag}
\usepackage{xcolor}
\usepackage[normalem]{ulem}
\usepackage{dsfont}
\usepackage{txfonts}
\usepackage{tikz}

\newcommand{\eeqa}{\end{eqnarray}}

\newcommand{\bra}[1]{\langle #1|}
\newcommand{\ket}[1]{|#1\rangle}

\newcommand{\para}[1]{\left( #1 \right)}

\def\beq{\begin{equation}}
\def\eeq{\end{equation}}
\def\beqa{\begin{eqnarray}}
\def\eeqa{\end{eqnarray}}
\def\bal#1\eal{\begin{align}#1\end{align}}
\def\bfig{\begin{figure}}
\def\efig{\end{figure}}

\newcommand{\eq}[1]{Eq.~(#1)}

\newcommand{\fig}[1]{Fig.~#1}

\begin{document}
\title{Holonomic quantum computation in the ultrastrong-coupling regime of circuit QED}
\author{Yimin Wang}
\affiliation{Quantum Physics and Quantum Information Division, \\
Beijing Computational Science Research Center, Beijing 100193, China}
\affiliation{College of Communications Engineering, PLA University of Science and Technology, Nanjing 210007, China}
\author{Jiang~Zhang}
\affiliation{Quantum Physics and Quantum Information Division, \\
Beijing Computational Science Research Center, Beijing 100193, China}
\author{Chunfeng Wu}
\affiliation{Pillar of Engineering Product Development, Singapore University
of Technology and Design, 8 Somapah Road, Singapore 487372}
\author{J.~Q.~You}
\email{jqyou@csrc.ac.cn}
\affiliation{Quantum Physics and Quantum Information Division, \\
Beijing Computational Science Research Center, Beijing 100193, China}
\author{G. Romero}
\affiliation{Departamento de F\'{\i}sica, Universidad de Santiago de Chile (USACH), Avenida Ecuador 3493, 9170124, Santiago, Chile}

\begin{abstract}
We present an experimentally feasible scheme to implement holonomic quantum computation in the ultrastrong-coupling regime of light-matter interaction. The large anharmonicity and the $\mathbb{Z}_2$ symmetry of the quantum Rabi model allow us to build an effective three-level $\Lambda$-structured artificial atom for quantum computation. The proposed physical implementation includes two gradiometric flux qubits and two microwave resonators where single-qubit gates are realized by a two-tone driving on one physical qubit, and a two-qubit gate is achieved with a time-dependent coupling between the field quadratures of both resonators. Our work paves the way for scalable holonomic quantum computation in ultrastrongly coupled systems.

\end{abstract}

\pacs{03.67.Lx, 85.25.-j, 42.50.Dv}

\maketitle

\section{Introduction}

%
%


The extensive progress in quantum information science has motivated continuous demand for implementing high-fidelity quantum operations. Holonomic quantum computation (HQC) represents a promising approach to achieve this goal because of its intrinsic noise-resilience features \cite{paolo1999,sjoqvist12}. The holonomic gates can be achieved by using either Abelian \cite{berry1984,aharonov87} or non-Abelian \cite{wilczek1984,anandan88} geometric phases. The Abelian approach \cite{jones1999,wang01,zhu02} utilizes quantum two-level systems (qubits) as elementary units, and the underlying idea is to choose a pair of orthogonal states that will evolve cyclically. In contrast, the non-Abelian approach \cite{paolo1999,duan01} embeds qubits in a proper subspace of the total Hilbert space. Recently, a scheme to build fast holonomic gates through the non-Abelian approach has been proposed in Refs.~\cite{sjoqvist12,xu12}. The advent of this idea has triggered off a set of new proposals \cite{zhang14,zhang15}. Apart from the theoretical interest, high fidelity gates based on fast HQC schemes have been demonstrated in different systems such as transmon-based superconducting qubits \cite{abdumalikov13}, NMR systems \cite{feng13}, and diamond NV-centers \cite{arroyo-camejo14,zu14}.

On the other hand, the light-matter interaction has been the focus of interest in recent years owing to the experimental realizations of the ultrastrong-coupling (USC) regime~\cite{ana2009,gunter2009,todorov2010,scalari2012,pol10,niemczyk10,chen16}. In this case, the light-matter coupling strength is comparable to the cavity and the qubit frequencies \cite{casanova2010}, and in the dipolar approximation, it is described by the quantum Rabi model (QRM)~\cite{rabi1936,braak2011}. Apart from the fundamental interest of the USC regime, it has been intensively studied for demonstrating novel quantum optics phenomena \cite{ashhab2010,ridolfo2012,sanchez-burillo2014,garziano2015,hwang2015}, implementing quantum information tasks \cite{kyaw2015,felicetti2015}, as well as fast quantum computation \cite{wang2004,wang2009,nataf2011,wang2012,romero2012} within circuit quantum electrodynamics (QED)~\cite{wallraff2004,chiorescu2004}. The latter provides a promising solid-state architecture for performing quantum computation due to the desirable properties of superconducting qubits, such as long coherence times, and most importantly, its controllability and scalability~\cite{barends2014}.

Meanwhile, many efforts have also been made to implement HQC in cavity QED system with natural atoms \cite{recati2012} and artificial atoms \cite{zhou2015}, as well as circuit QED system with superconducing qubits \cite{xue2016}. While the HQC scheme in Ref.~\cite{recati2012} is performed adiabatically, the schemes in Refs.~\cite{zhou2015,xue2016} are designed in a non-adiabatic fashion in addition to their decoherence-free subspace (DFS) encoding, and thus integrate both the noise resilience of DFS and the operational robustness of holonomies. However, all of those schemes \cite{recati2012,zhou2015,xue2016} are based on the strong coupling of light-matter interaction, which can be well described by the Jaynes-Cummings model. To the best of knowledge, so far no scheme has ever been proposed to construct holonomic gates in the ultrastrong-coupling regime with the quantum Rabi model. Ultrastrong coupling offers the possibility for ultrafast quantum gate operations even in the time scale of subnanosecond \cite{romero2012}, therefore, the realization of HQC in USC is of particular interest yet challenging.

In this work, we propose an experimentally feasible scheme to implement universal non-adiabatic HQC in the USC regime of light-matter interaction. The large anharmonicity and the $\mathbb{Z}_2$ symmetry of the QRM allow us to construct an effective three-level $\Lambda$-type system for quantum computing. We show that non-commuting single-qubit holonomic gates can be obtained by means of a two-tone driving on one physical qubit, and nontrivial two-qubit holonomic gates can be achieved with a time-dependent coupling between the field quadratures of two bosonic modes. Moreover, we discuss the physical implementation by considering two gradiometric flux qubits each galvanically coupled to its transmission line resonator, which are then connected to each other through a superconducting quantum interference device.
Compared to the existing proposals for implementing HQC in circuit QED, the strategy we pursue is different in the sense that we exploit the discrete $\mathbb{Z}_2$ symmetry of the quantum Rabi model instead of the continuous $U(1)$ symmetry of the Jaynes-Cummings model. Therefore, our proposal works well in the ultrastrong-coupling regime, and it may find compelling applications for quantum information processing in ultrastrong coupling and deep strong coupling regimes for various systems.

\section{Selection rules in the Quantum Rabi model}
\label{sec_rabi}
\begin{figure}[b]
\centering
\includegraphics[scale=0.3]{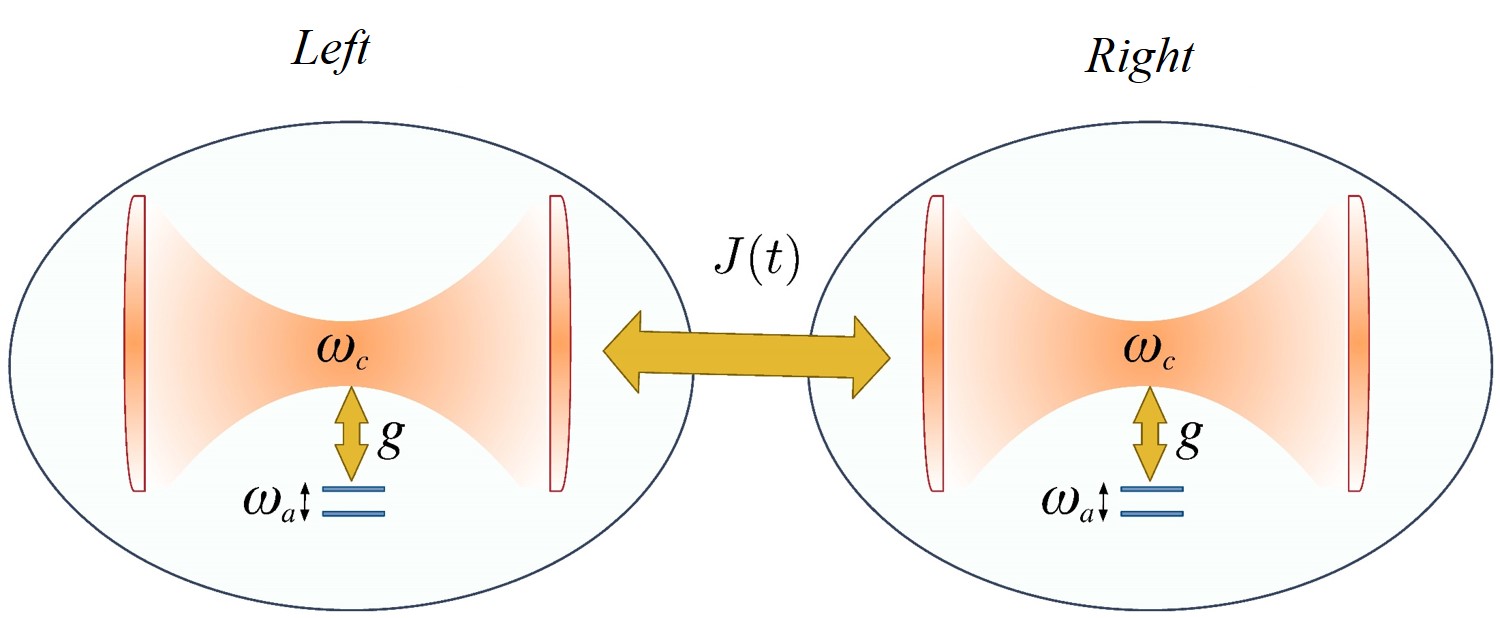}
\caption{(Color online) Schematic representation of our model. A system of a single qubit and a single cavity mode that interact in the ulstrastrong-coupling regime constitutes the quantum Rabi system. The interaction between the two quantum Rabi systems is mediated by cavities through a time-dependent coupling of strength $J(t)$.}
\label{fig1}
\end{figure}
The model that we consider is schematically depicted in Fig.~\ref{fig1}. It includes two ultrastrongly coupled qubit-cavity systems, which interact via a time-dependent coupling of strength $J(t)$. Each ultrastrongly coupled system, onwards called quantum Rabi system (QRS), is described by
\beq
H_p = \hbar \omega_c a^\dagger a +  \hbar  \frac{\omega_{a}}{2} \sigma_{z} +  \hbar g \sigma_{x} (a^\dagger + a),
\label{eq_rabi1}
\eeq
where $\omega_{a}, \omega_c$, and $g$ stand for the qubit frequency, cavity frequency, and the qubit-resonator coupling strength, respectively. In addition, $a(a^\dagger)$ is the bosonic annihilation(creation) operator, and $\sigma_{z},\, \sigma_{x}$ are the Pauli matrices of the qubit.
\begin{figure}[t]
\begin{center}
\includegraphics[scale=0.4]{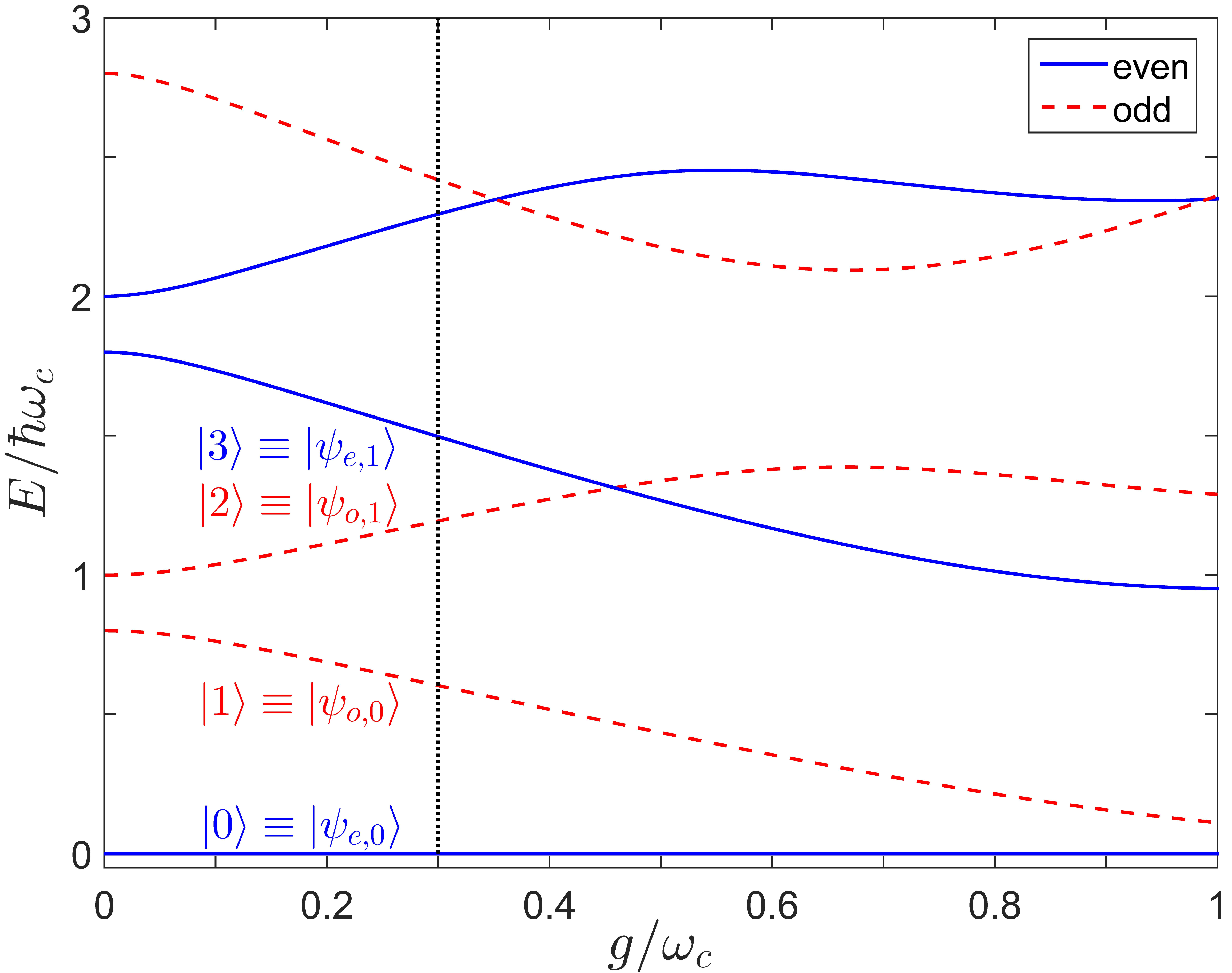}
\caption{Energy levels of the quantum Rabi model as a function of the dimensionless parameter $g/\omega_c$ with $\omega_a/\omega_c=0.8$. Energies are rescaled in order to set the ground level to zero.
The parity of the corresponding eigenstates is identified, continuous-blue line for even states and dashed-red lines for odd states. }
\label{fig2}
\end{center}
\end{figure}

In the ultrastrong-coupling regime~\cite{ciuti2005,bourassa2009}, which is characterized by the ratio range $0.1 \lesssim g/\omega_c < 1$, the bosonic field and the qubit merge into dressed-state systems that feature the discrete $\mathbb{Z}_2$ symmetry, as shown in \fig{\ref{fig2}}, where we plot the energy spectrum of the quantum Rabi model as a function of the coupling strength $g/\omega_c$.
This symmetry is characterized by the parity operator $\mathcal{P} = e^{i\pi (a^\dagger a + \sigma_+\sigma_-)}$, such that $\mathcal{P} \ket{\psi_e}= \ket{\psi_e}$, $\mathcal{P} \ket{\psi_o}=- \ket{\psi_o}$ . Note that in \fig{\ref{fig2}}, even$(\ket{\psi_e})$ and odd$(\ket{\psi_o})$ eigenstates are represented by continuous-blue and dashed-red lines, respectively.
The QRM can be rewritten as
\bal
H_p=& \sum_{s=0}\hbar \omega_{s}\ket{s}\bra{s},
\label{eq_rabi3}
\eal
where we consider both even and odd parity states together and labeled them as eigenstates $\ket{{s}}$ of increasing energy $\hbar\omega_{s}$.

Formally, the parity in quantum mechanics is intimately related to the selection rules. For the QRM It can be shown that the matrix elements of an even operator are zero between states of different parity, $\bra{\psi_e}\mathcal{A}_e \ket{\psi_o}=\bra{\psi_o}\mathcal{A}_e\ket{\psi_e}=0$, while the matrix elements of an odd operator are zero between states of equal parity $\bra{\psi_e}\mathcal{A}_o \ket{\psi_e}=\bra{\psi_o}\mathcal{A}_o \ket{\psi_o}=0$. Also, from \fig{\ref{fig2}}, we see that the spectrum is anharmonic enough such that the dressed states may be used as computational basis for quantum information processing.
In particular, when $g/\omega_c = 0.3$, one can build an effective three-level system by choosing the lowest three levels, $\ket{0} \equiv \ket {\psi_{e,0}}$, $\ket{1} \equiv \ket {\psi_{o,0}}$ and $\ket{2} \equiv \ket {\psi_{o,1}}$ to implement holonomic quantum computation schemes.
\section{Single-qubit gate}
In this section we show how to construct an arbitrary single-qubit gate in the dressed-state basis of the quantum Rabi model with a non-adiabatic non-Abelian scheme \cite{sjoqvist12}.  We choose the two lower levels $\ket{0}$ and $\ket{1}$ to form the qubit subspace $\mathcal{S}_1(0)\equiv \{\ket{0},\ket{1}\}$, leaving the upper level $\ket{2}$ as an auxiliary state.

In this encoding, the states $\ket{0}$ and $\ket{2}$ belong to different parity subspaces, such that the transitions between them can be induced by an odd parity operator, i.e. $\sigma_x$. Similarly, the states $\ket{1}$ and $\ket{2}$ have the same parity and the transition between them can be induced by an even parity operator such as $\sigma_z$. Therefore, a single-qubit holomonic quantum gate can be realized by making use of a two-tone driving scheme on the physical qubit. This can be modeled by the Hamiltonian
\bal
H_d=& \Omega_1(t) \cos (\bar{\omega}_1 t + \varphi_1) \sigma_x + \Omega_2(t) \cos (\bar{\omega}_2 t + \varphi_2) \sigma_z.
\label{eq_1rabi2qd0}
\eal

The qubit driving Hamiltonian \eq{\ref{eq_1rabi2qd0}} can be written in the dressed-state basis by using the completeness relation $I = \sum_s \ket{s}\bra{s}$,
\bal
H_d=& \Omega_1(t) \cos (\bar{\omega}_1 t + \varphi_1) \sum_{s, t} \,\,x_{st}\ket{s}\bra{t} \nonumber \\
&+ \Omega_2(t) \cos (\bar{\omega}_2 t + \varphi_2) \sum_{s,t} \,\,z_{st}\ket{s}\bra{t} ,
\label{eq_1rabi2qd1}
\eal
where the transition elements are given by $x_{st} = \bra{s}\sigma_x\ket{t}$ and $z_{st} = \bra{s}\sigma_z\ket{t}$. Notice that according to the selection rule for even and odd operators, $x_{st} =0$ if $\ket{s}$ and $\ket{t}$ belong to same parity subspace, and $z_{st}=0$  if $\ket{s}$ and $\ket{t}$ belong to different parity subspace. Furthermore, we can interpret the projector $\ket{s}\bra{t}$ as a flip operator between dressed states of either  equal or different parity depending on the nature of the system operator (in our case, either $\sigma_x$ or $\sigma_z$). Therefore, such a Hamiltonian \eq{\ref{eq_1rabi2qd1}} induces coherent excitation transfer between all the possible dressed states.
\bfig[b]
\begin{center}
\includegraphics[scale=0.44]{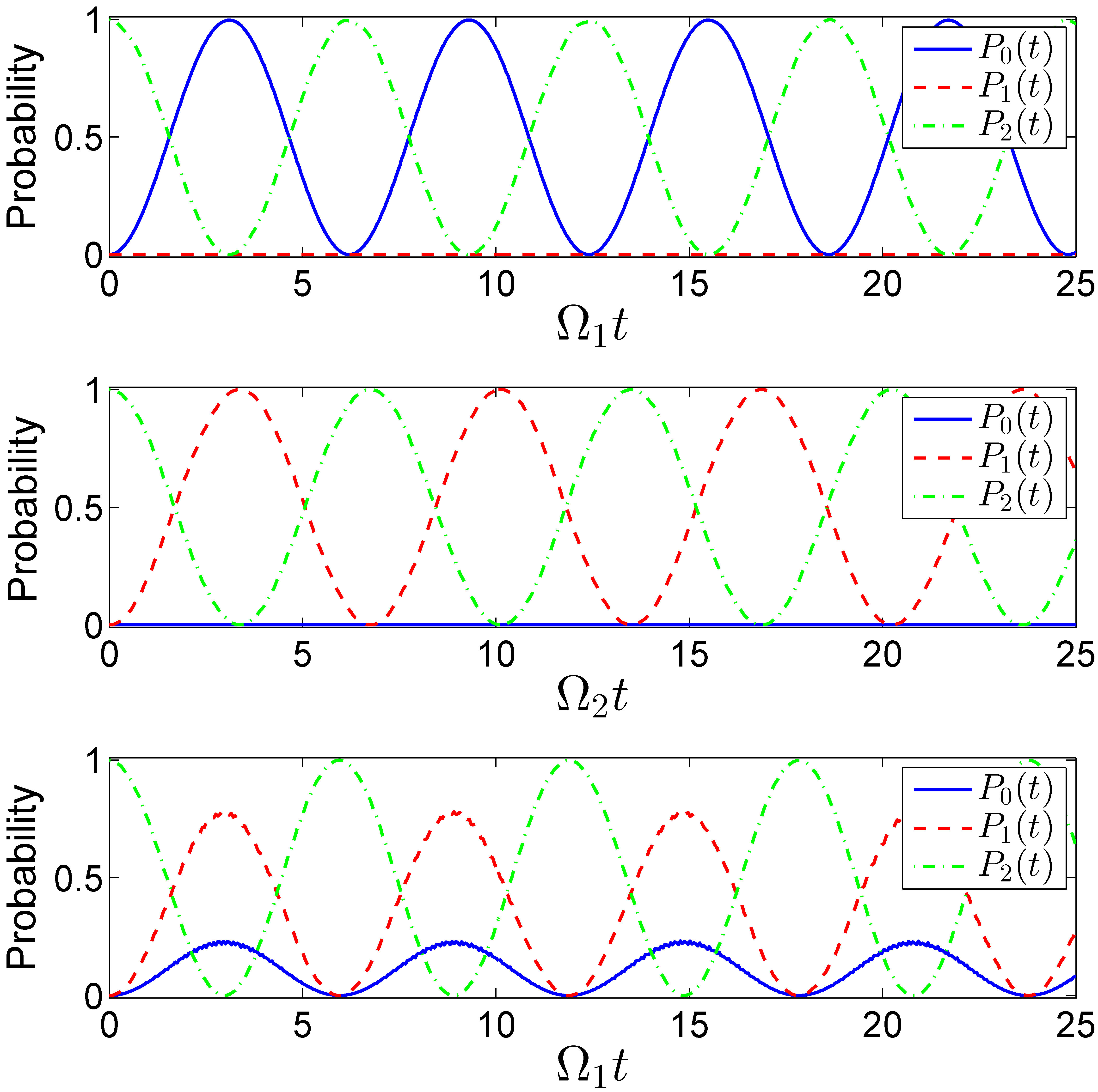}
\caption{Rabi oscillations for the lowest three dressed states in the quantum Rabi model by driving the physical qubit: (a) only in the ${\bf x}$ direction with strength $\Omega_1 = 0.02  \omega_{20}$, $\Omega_2 = 0$; (b) only in the ${\bf z}$ direction with strength $\Omega_1 = 0 $, $\Omega_2 = 0.02  \omega_{21} $; (c) both in the ${\bf z}$ and ${\bf x}$ directions with strengthes $\Omega_1 = \Omega_2 = 0.02  \omega_{21}$. $P_i(t)\, (i=0,1,2)$ is the probability of the state $\ket{i}$ as a function of time. The parameters are $\omega_a=0.8\omega_c$, $g=0.3\omega_c$, and driving frequencies $\bar{\omega}_1 =\omega_{21}$, $\bar{\omega}_2 =\omega_{20}$.}
\label{fig3}
\end{center}
\efig

As we have shown in \fig{\ref{fig2}}, for $g/\omega_c \leq 1$, the energy spectrum of the quantum Rabi system has a large anharmonicity such that one can access one particular transition frequency $\omega_{ts} =\omega_{t} - \omega_{s}$. Let us consider the total Hamiltonian $H = H_p + H_d$, and we move to the interaction picture with respect to the Rabi Hamiltonian in \eq{\ref{eq_rabi3}}. If the condition $|\Omega_j(t)|\ll \min\{|\omega_{t,s}|,\bar{\omega}_j\}$ is satisfied, one can apply the RWA and neglect fast oscillating terms. Moreover, when bringing the frequency of the driving close to resonance with the transitions in which we are interested, i.e., $\bar{\omega}_1 = \omega_{20}$ and $\bar{\omega}_2 = \omega_{21}$, the interaction Hamiltonian reads
\bal
H_I=& \frac{\Omega_1(t)}{2} e^{-i \varphi_1} \,x_{20} \ket{2}\bra{0} + \frac{\Omega_2(t)}{2} e^{-i \varphi_2} \, z_{21} \ket{2}\bra{1} + {\rm H.c.},
\label{eq_1rabi2qd2}
\eal
with $x_{20} = \bra{2}\sigma_x\ket{0}$ and $z_{21} = \bra{2}\sigma_z\ket{1}$. Therefore, by engineering the driving amplitudes and frequencies, Rabi oscillations between two specific dressed states can be performed.

In \fig{\ref{fig3}}, we show Rabi oscillations for the lowest three dressed states in the quantum Rabi model. This simulation has been performed by making use of the full Hamiltonian $H=H_p+H_d$. As shown in \fig{\ref{fig3}} (a), by driving the qubit in the $\sigma_x$ direction on resonance with the transition $\bar{\omega}_1 =\omega_{20}$, we observe Rabi oscillations between the two different parity states $\ket{2}$ and $\ket{0}$. Similarly, by driving the qubit in the $\sigma_z$ direction on resonance with the transition $\bar{\omega}_2 =\omega_{21}$, the complete population transfer between the two same parity states $\ket{2}$ and $\ket{1}$ is shown in \fig{\ref{fig3}} (b). Moreover, by tuning the driving frequency and amplitude to make both $\sigma_x$ and $\sigma_z$ rotations, we may have full control of any structured three-level system built from the dressed states of the quantum Rabi model, as shown in \fig{\ref{fig3}} (c). In this manner, we are able to implement the effective Hamiltonian in \eq{\ref{eq_1rabi2qd2}}.

Now we show how to construct an arbitrary holonomic single-qubit gate with the effective Hamiltonian $H_I$. By setting $\Omega^2(t) = (\Omega^2_1(t) {|x_{20}|}^2 + \Omega^2_2(t){|z_{21}|}^2)/4$, $\varphi=\varphi_2-\varphi_1$, and  $\theta = -2 \arctan[(\Omega_1(t)x_{20})/(\Omega_2(t)z_{21})]$, we can rewrite \eq{\ref{eq_1rabi2qd2}} as follows
\beq
\label{eq_lambdahi}
H_I=\Omega(t)\left(e^{i\varphi}\sin\frac{\theta}{2}\,\ket{2}\bra{0}-\cos \frac{\theta}{2}\,\ket{2}\bra{1}+ {\rm H.c.}\right).
\eeq
In this case, we construct a $\Lambda$-system Hamiltonian in the dressed-state basis, from which an arbitrary single-qubit holonomic gate can be obtained. The effective Hamiltonian $H_I$ in \eq{\ref{eq_lambdahi}} can be recast as the auxiliary state $\ket{2}$ coupled to the bright state $\ket{b}=e^{-i\varphi}\sin\frac{\theta}{2}\,\ket{0}-\cos \frac{\theta}{2}\,\ket{1}$ and decoupled from the dark state $\ket{d}=e^{i\varphi}\sin\frac{\theta}{2}\,\ket{1}+\cos \frac{\theta}{2}\,\ket{{0}}$. Initially, quantum information is stored in the qubit states of subspace $\mathcal{S}_1(0)$. When $H_I$ is applied, the subspace $\mathcal{S}_1$ is driven out of $\mathcal{S}_1(0)$, and we obtain the Rabi oscillation between states $\ket{b}$ and $\ket{{2}}$ with Rabi frequency of $\Omega(t)$. When the condition $\int_0^{\tau}\Omega(t)\mathbf{d}t=\pi$ is satisfied, the system states return to the original subspace $\mathcal{S}_1(0)$ after a cyclic evolution. The corresponding unitary operator $U_I(\tau)$ acting on $\mathcal{S}_1(0)$ reads $U_I(\tau)=\mathbf{n}\cdot\mathbf{\sigma}$, where $\mathbf{n}=(\sin\theta\cos\varphi, \sin\theta\sin\varphi, \cos\theta)$ and $\mathbf{\sigma}=(\sigma_x, \sigma_y, \sigma_z)$ being Pauli operators~\cite{sjoqvist12}. It is clear that two non-commuting single-qubit holonomic quantum gates can be achieved based on $U_I(\tau)$. Moreover, there is no dynamical contribution to $U_I(\tau)$ since $\bra{{m}}H_I\ket{{n}}=0$ $(m,n\in \mathds{N})$ and hence $\bra{{m}(t)}H_I\ket{{n}(t)}=0$. The result shows the pure geometric nature of the obtained gate. Therefore, the desired single-qubit gates for universal quantum computation can be implemented in our system based on the non-adiabatic non-Abelian scheme \cite{sjoqvist12,xu12}.

The holonomic single-qubit gate performance under loss mechanisms in the ultrastrongly coupled system can be studied by means of the time-convolutionless projection operator method~\cite{QOSBook}. In this approach the master equation reads
\begin{equation}
\dot{\rho}=\frac{1}{i\hbar}[H_s,\rho]+\sum_{n=z,x,c}(U_n\rho S_n+S_n\rho U^{\dag}_n-S_nU_n\rho-\rho U^{\dag}_nS_n),
\label{TCPOM}
\end{equation}
where $S_n$ are Hermitian system operators, and the operators $U_n$ are defined as
\begin{align}
U_n =&\int_0^{\infty}d\tau~\nu_n(\tau)e^{-(i/\hbar) H_s\tau}S_ne^{(i/\hbar) H_s\tau}\nonumber\\
\nu_n(\tau)=&\int_{-\infty}^{\infty}d\omega~\frac{\gamma_n(\omega)}{2\pi}[\bar{N}_n(\omega)e^{i\omega\tau}+(\bar{N}_n(\omega)+1)e^{-i\omega\tau}].
\end{align}
Here, we consider independent thermal baths for each loss mechanism acting on the system described by the Hamiltonian $H_s(t)=\Omega(t)(\Upsilon_0\ket{2}\bra{0})+\Upsilon_1\ket{2}\bra{1}+{\rm H.c.})$ [cf. Eq.~(\ref{eq_lambdahi})]. In our simulation we include loss mechanisms acting on the dressed-state system via transversal noise ($\gamma_x$), longitudinal noise ($\gamma_z$), and noise acting on the field quadrature ($\gamma_c$), through operators $S_{x}=\sigma_x$, $S_{z}=\sigma_z$, and $S_c=a + a^{\dag}$, respectively. In our treatment, each loss mechanism is described by independent thermal baths with bare loss rates $\gamma_j$. This leads to $\gamma_j(\omega)=(\gamma_j/\omega_j)\omega\Theta(\omega)$, where $\Theta(\omega)$ is the Heaviside step function.

Following Ref.~\cite{sjoqvist12} we have studied the performance of the Hadamard gate under loss mechanisms through the gate fidelity $F=\langle\chi| U^{\dag}(C)\rho_{\rm out}U(C)|\chi\rangle$, where $U(C)=(\sigma_x+\sigma_z)/\sqrt{2}$, and $\rho_{\rm out}$ is the density matrix of the output state obtained from the master equation~(\ref{TCPOM}). The gate fidelity is computed numerically for $4000$ input states $|\chi\rangle$, uniformly distributed over the Bloch sphere. In $H_s(t)$, we choose $\Omega(t)(\Upsilon_0,\Upsilon_1)=\beta {\rm sech}(\beta t)(1,(\sqrt{2}-1))/\sqrt{2(2-\sqrt{2})}$, and the pulse is truncated where the amplitude is $\beta/1000$, which gives the pulse with a length $\tau=(2/\beta) {\rm arcsech}(1/1000)$. The result is shown in Fig.~\ref{fig4} for parameters $g=0.3\omega_c$, $\omega_a=0.8\omega_c$, and $\gamma_x=\gamma_z=\gamma_c=10^{-2}\omega_c$. Note that loss mechanisms in the dressed-state basis, including even and odd operators in the parity Hilbert space, will induce a complete decay to the fundamental state $\ket{0}$ at a scale time of $\sim 100/\omega_c$. Despite of this, if the pulse is sufficiently short compared with the decay time, i.e. $\beta/\gamma_x\gg1$, the fidelity of the nonadiabatic gate approaches to unity.
\bfig[h!]
\begin{center}
\includegraphics[scale=0.44]{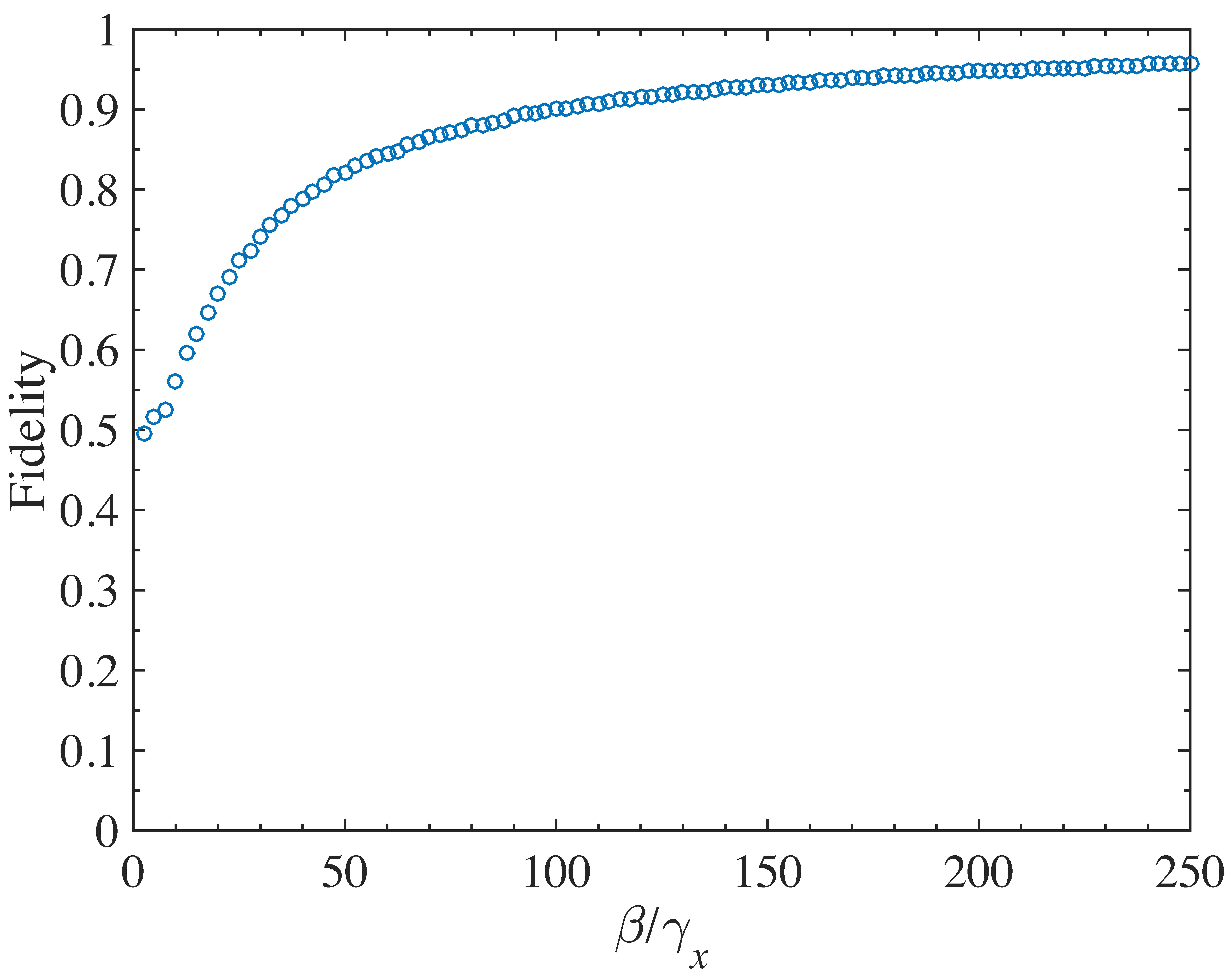}
\caption{Performance of the Hadamard gate under loss mechanisms acting upon the ultrastrongly coupled system. The fidelity is averaged on an ensemble of $4000$ input states uniformly distributed over the Bloch sphere. In this simulation we have considered parameters $g=0.3\omega_c$, $\omega_a=0.8\omega_c$, $\gamma_x=\gamma_z=\gamma_c=10^{-2}\omega_c$.}
\label{fig4}
\end{center}
\efig

\section{Two-qubit gate}
In what follows, we will demonstrate a nontrivial two-qubit gate by using a non-adiabatic Abelian scheme \cite{zhu02} in a four-dimensional space spanned by the encoded logic qubit states $\mathcal{S}_2\equiv \{\ket{{0}_l{0}_r},\ket{{1}_l{0}_r}, \ket{{0}_l{1}_r}, \ket{{1}_l{1}_r}\}$. This can be proven by considering two ultrastrongly coupled systems that interact via a time-dependent coupling strength $J(t)$, as depicted in \fig{\ref{fig1}}. The Hamiltonian describing the whole system is composed of the sum of two quantum Rabi models and a coupling between the field quadratures~\cite{felicetti2014,rossatto2015}
\bal
\label{eq_fullh1}
H_{\rm tot} & = H_{p,l} + H_{p,r} + H_{\rm int}, \nonumber \\
H_{\rm int} &=\hbar J(t) ( a_l^\dagger + a_l) (a_r^\dagger + a_r ),
\eal
with $H_{p,j}\,\,(j = l,r)$ being the Hamiltonian for the left and right quantum Rabi system.

By using the completeness relation, the system Hamiltonian \eq{\ref{eq_fullh1}} can be rewritten as
\bal
H_{\rm tot} =& \sum_{{s}=0}  \big(\hbar\omega_{{s},l} \ket{{s}_l}\bra{{s}_l}  + \hbar\omega_{{s},r} \ket{{s}_r}\bra{{s}_r} \big) + \hbar J(t)\times \nonumber \\
&\Big[\sum_{{s}_l,{t}_l>{s}_l}\big(f_{{s}_l{t}_l}\ket{{s}_l}\bra{{t}_l} + {\rm H.c.} \big)\otimes \sum_{{u}_r,{v}_r>{u}_r}\big(f_{{u}_r{v}_r}\ket{{u}_r}\bra{{v}_r}  + {\rm H.c.}\big)\Big],
\label{eq_2rabi1}
\eal
where $f_{{s}_j{t}_j} = \bra{{s}_j}(a_j + a_j^\dagger)\ket{{t}_j}$, $j = l,r$, is the transition matrix elements for the left ($l$) and right ($r$) system. Here, we have used the fact that the transition matrix elements are zero between states of the same parity, i.e., $f_{{s}_j{s}_j} =0$. Similar to the single-qubit case, the operator $\ket{{s}_j}\bra{{t}_j}$ is the raising operator for the left or the right system.
Let us consider the interaction picture with respect to Hamiltonian $H_{p,l} + H_{p,r}$. In this case, the interaction Hamiltonian reads
\bal
H^I_{\rm int} =& \hbar J(t)\Big[\sum_{{s_l},{t_l}>{s_l}}\big(f_{{s}_l{t}_l}\ket{{s}_l}\bra{{t}_l} e^{-i\omega_{{t}_l{s}_l} t} + {\rm H.c.} \big) \nonumber \\
&\times \sum_{{u_r},{v_r}>{u_r}}\big(f_{{u}_r{v}_r}\ket{{u}_r}\bra{{v}_r}e^{-i\omega_{{v}_r{u}_r} t}  + {\rm H.c.}\big)\Big],
\label{eq_2rabiip1}
\eal
where $\omega_{{t}_j{s}_j} =\omega_{{t},j}-\omega_{{s},j} >0$.
In particular, the cavity-cavity coupling parameter can be a time-dependent function $J(t)=J_0(t) \cos(\omega_d t + \varphi_d)$. In this case, if one chooses the resonance condition for two specific dressed states, i.e. $\omega_d = \omega_{{v}_r{u}_r}-\omega_{{t}_l{s}_l}$ and the cavity-cavity coupling strength satisfies the condition $|J_0(t)|\ll \omega_{{t}_l{s}_l}+ \omega_{{v}_r{u}_r}$, we can apply the rotating-wave approximation and the interaction Hamiltonian effectively reduces to
\bal
H_{\rm int} = \hbar \frac{J_0(t)}{2} f_{{s}_l{t}_l} f_{{u}_r{v}_r}^* e^{-i \varphi_d} \ket{{s}_l}\bra{{t}_l} \otimes\ket{{v}_r}\bra{{u}_r} + {\rm H.c.}.
\label{eq_2rabi2}
\eal
It is clear that such a Hamiltonian produces entanglement and induces coherent excitation transfer between specific dressed states $\ket{{s}_l}$ and $\ket{{t}_l}$ of the left and the right systems~\cite{rossatto2015}.
It is worthy noting that the coupling operator $(a_l^\dagger + a_l) (a_r^\dagger + a_r )$ is an odd operator for the left and right quantum Rabi system individually, so it only induces transitions between states with opposite parity.
Specifically, if we choose $\omega_d = \omega_{{1}_r{0}_r}-\omega_{{1}_l{0}_l}$, the system can be effectively described
\bal
H_{\rm int} = \hbar \frac{J_0(t)}{2} f_{{0}_l{1}_l} f^*_{{0}_r{1}_r} e^{-i \varphi_d}  \ket{{0}_l {1}_r}\bra{{1}_l {0}_r} + {\rm H.c.},
\label{eq_2rabi3}
\eal
which gives the interaction between two specific states $ \ket{{0}_l {1}_r}$ and $\ket{{1}_l {0}_r}$. Meanwhile, \eq{\ref{eq_2rabi3}} is our target Hamiltonian for the two-qubit HQC with the lowest three states in the dressed state basis. The interaction Hamiltonian~(\ref{eq_2rabi3}) also indicates that, out of the nine possible tensor states in the Hilbert space of the total system $\mathcal{H}=\mathcal{H}_{p,l}\otimes \mathcal{H}_{p,r}$, there are only two states that are correlated. This restricts us to a two-dimensional subspace which is spanned by vectors $\{\ket{{1}_l {0}_r}, \ket{{0}_l {1}_r}\}$, and allows us to demonstrate a two-qubit gate based on Abelian geometric phases. To illustrate our scheme, we encode the states $\ket{{1}_l {0}_r}$ and $\ket{{0}_l {1}_r}$ into logical single-qubit states $\ket{1}_L$ and $\ket{0}_L$, respectively. In the logical representation, \eq{\ref{eq_2rabi3}} reads
\begin{equation}
\label{eq_hint}
H_{\rm int}= \hbar \frac{J_0(t)}{2} f_{{0}_l{1}_l} f^*_{{0}_r{1}_r} (\cos \varphi_d \, S_x + \sin \varphi_d \, S_y),
\end{equation}
where $S_x$ and $S_y$ are Pauli operators on the logical basis.

In what follows, we demonstrate a nontrivial two-qubit gate based on the Hamiltonian \eq{\ref{eq_hint}} according to the non-adiabatic Abelian scheme presented in Ref.~\cite{zhu02}. A geometric phase shift gate $U_{\beta}$ acting on the eigenstates of $S_x$, namely $\ket{\pm}_L=(\ket{0}_L\pm\ket{1}_L)/\sqrt{2}$, can be achieved by letting the system evolve along a cyclic path based on properly designed four-step evolution up to a global phase. {\it Step-1.---} By setting $\varphi_d=\pi/2$, we apply a rotation $e^{i \pi S_y/4}$ to the basis, changing the states from $\ket{+}_L$ and $\ket{-}_L$ to $\ket{0}_L$ and $-\ket{1}_L$, respectively. {\it Step-2.---} By setting $\varphi_d=0$, the states $\ket{0}_L$ and $\ket{1}_L$ are swapped with each other by a rotation $e^{i \pi S_x/2}$. {\it Step-3.---} With a proper choice of the parameter $\varphi_d = \beta \neq m \pi$ ($m\in\mathds{N}$), we evolve the states from $\ket{0}_L$ and $\ket{1}_L$ to $e^{i\beta}\ket{1}_L$ and $e^{-i\beta}\ket{0}_L$ respectively, by a rotation $e^{i \pi(\cos\beta S_x+\sin\beta S_y)/2}$.  {\it Step-4.---} By choosing $\varphi_d =\pi/2$ again, the resulting states $\ket{0}_L$ and $-\ket{1}_L$ can be changed back to $-\ket{+}_L$ and $\ket{-}_L$ by using the rotation $e^{-i\pi S_y/4}$, eliminating the minus sign obtained in the first step. Therefore, the system undergoes a cyclic evolution
\bal
&\ket{+}_L \rightarrow  \ket{0}_L \rightarrow   \ket{1}_L \rightarrow e^{-i \beta }\ket{0}_L  \rightarrow e^{-i \beta }\ket{+}_L, \\
&\ket{-}_L \rightarrow -\ket{1}_L  \rightarrow  -\ket{0}_L \rightarrow  -e^{i \beta }\ket{1}_L  \rightarrow  e^{i \beta }\ket{-}_L,
\eal
and the obtained geometric phase shift gate $U_{\beta}$ written in the states $\ket{\pm}_L$ is of the following form,
\begin{eqnarray}
 U_{\beta}=\left(
       \begin{array}{cc}
         e^{-i\beta} & 0 \\
         0 & e^{i\beta} \\
       \end{array}
     \right).
\label{eq_ubeta1}
\end{eqnarray}
There is no dynamical phase accompanied during the cyclic evolution since the evolution is along geodesic lines. \eq{\ref{eq_ubeta1}} is nothing but a non-trivial two-qubit gate in the basis $\{\ket{{0}_l{0}_r},\ket{{1}_l{0}_r}, \ket{{0}_l{1}_r}, \ket{{1}_l{1}_r}\}$ with
\beq
U_2=\left(
     \begin{array}{cccc}
       1 & 0 & 0 & 0 \\
       0 & \cos\beta & -i\sin\beta & 0 \\
       0 & -i\sin\beta & \cos\beta & 0 \\
       0 & 0 & 0 & 1 \\
     \end{array}
   \right).
\eeq
It is apparent that $U_2$ is nontrivial when $\beta\neq m\pi$ with $m\in\mathds{N}$. Together with the non-commuting single-qubit gates, we have demonstrated a universal set of holonomic quantum gates for ultrastrongly coupled system.
\section{Physical Implementation}
\label{sec phyimp}
\bfig[b]
\begin{center}
\includegraphics[scale=0.5]{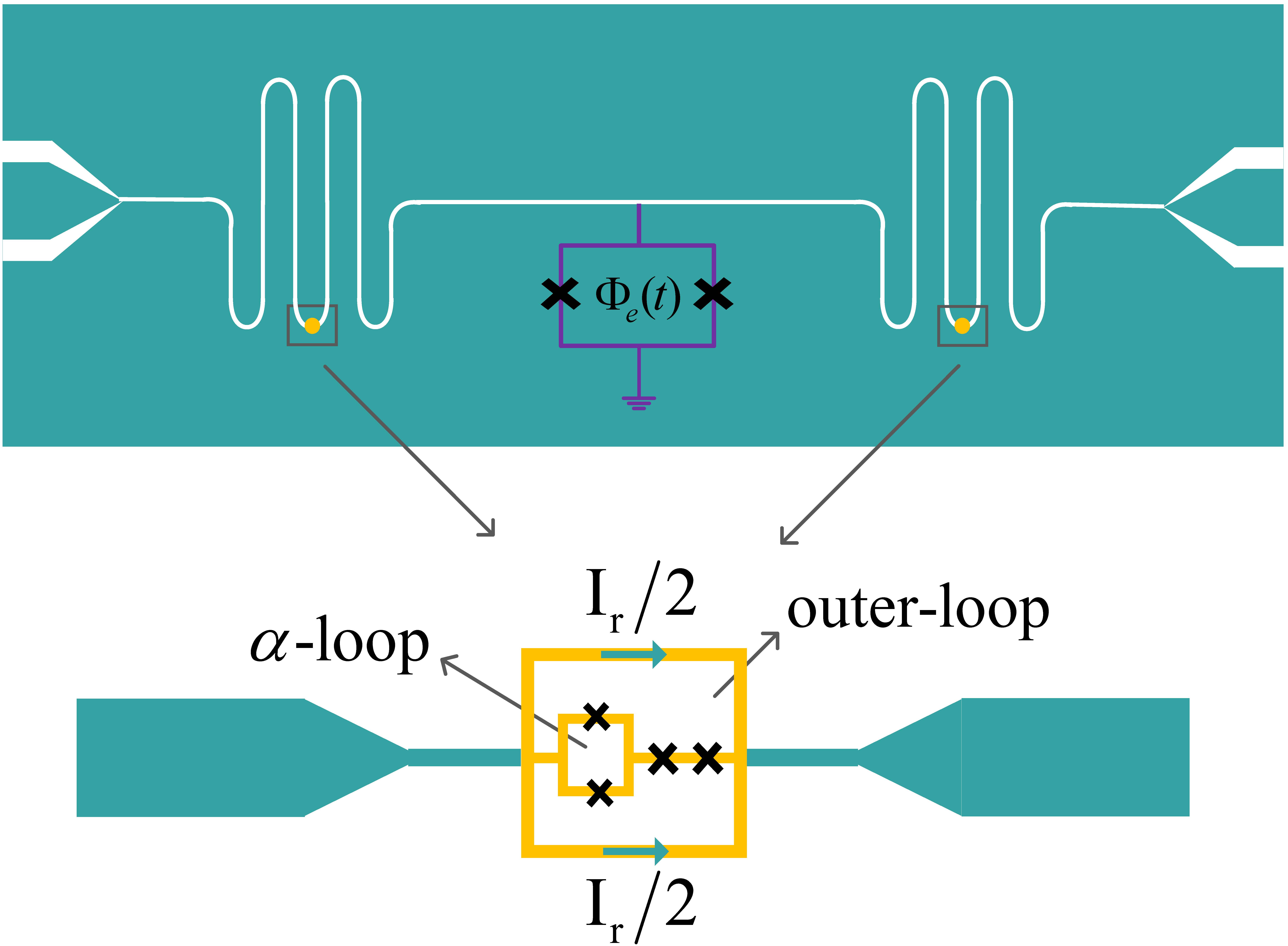}
\caption{Schematic of circuit QED design for the holonomic quantum computation. Two transmission line resonators (cavities) are grounded through a SQUID. Each cavity is galvanically coupled to a gradiometric tunable-gap flux qubit, that is constituted by four Josephson junctions. The time-dependent interaction between two resonators can be realized by modulating the external magnetic flux $\Phi_e(t)$ through the SQUID.}
\label{fig5}
\end{center}
\efig
Here, we propose the use of a gradiometric flux qubit with a tunable gap galvanically connected to a cavity, see Fig.\ref{fig5}, to implement HQC. This circuit QED architecture in the strong-coupling regime has been implemented in~\cite{schwarz2013}. Also, the ultrastrong-coupling regime may be achieved by implementing a longer and thinner shared line between the gradiometric flux qubit and the microwave resonator~\cite{schwarz2015}.

We stress that the gradiometric configuration is unaffected by homogenous magnetic fields, so it has the advantage to overcome flux crosstalk~\cite{schwarz2013}. Also, an inhomogenous magnetic field in the outer-loop of the flux qubit enables the coupling to a microwave resonator. The additional $\alpha$-loop in the gradiometric configuration allows for a tunable qubit gap. This mechanism is completely independent of the flux line that controls the frustration parameter in the outer loop~\cite{schwarz2013,schwarz2015}. Therefore, our two-tone driving scheme for the single-qubit gate may be implemented in the gradiometric qubit by applying two independent magnetic fluxes of different frequencies to the outer loop and the $\alpha$-loop~\cite{wang2009}.

The time-dependent coupling $J(t)$ between the two cavities can be implemented by means of a superconducting quantum interference device (SQUID) \cite{felicetti2014}, threaded by an external flux $\Phi_e(t)$, as shown in \fig{\ref{fig5}}. Although the effective cavity length is oscillating with small deviations from its average value, we can still consider the system as a single-mode resonator, see Ref.~\cite{felicetti2014} for a detailed discussion. In particular, the specific form of cavity-cavity coupling strength ($J(t)=J_0(t) \cos(\omega_d t + \varphi_d)$) required for the two-qubit gate, may be achieved by choosing the time-dependent external magnetic flux $\Phi_e(t)$ to be composed of the sum of a small amplitude-modulated signal oscillating at the driving frequency $\omega_d$ and a constant offset $\bar\Phi$, namely, $\Phi_e(t) = \bar\Phi + \Delta\Phi \,\Omega(t) \cos{\para{\omega_d t + \varphi_d}}$. By controlling the driving frequency $\omega_d$, it will allow us to selectively activate the interaction between two specific energy states of the system and to obtain the effective Hamiltonian \eq{\ref{eq_2rabi3}} for the two-qubit gate, see the Appendix for a detailed discussion.
\section{Conclusion}
\label{sec Conclusion}
In conclusion, we have presented a proposal to implement a holonomic quantum computation scheme in the ultrastrong-coupling regime of circuit QED. The effective three-level $\Lambda$ artificial atom to carry out the quantum gate operations is built from the eigenstates of the quantum Rabi model in the dressed-state basis, which is based on its large anharmonicity and the $\mathbb{Z}_2$ symmetry. Arbitrary non-Abelian single-qubit gates can be achieved by selectively driving the physical qubit in both $\sigma_x$ and $\sigma_z$ directions with different frequencies. A non-trivial Abelian two-qubit quantum phase gate can be demonstrated by controlling both the frequency and the amplitude of the time-dependent cavity-cavity coupling strength $J(t)=J_0(t) \cos(\omega_d t + \varphi_d)$ between the field quadratures, and by performing a four-step cyclic evolution scheme. Furthermore, we have proposed an experimentally feasible circuit QED architecture for the physical implementation of our scheme. Here, each gradiometric tunable-gap flux qubit is connected galvanically to the center conductor of a transmission line resonator to reach the USC regime. The resonators are connected to the same edge of a grounded SQUID, which is surrounded by a time-dependent external magnetic flux. Our proposal provides novel applications of the ultrastrong-coupling regime of light-matter for scalable holonomic quantum computation.

\section{Acknowledgements}

This work was supported by National Basic Research Program of China Grant Nos. 2016YFA0301200 and 2014CB921401, NSAF Grant Nos. U1330201 and  U1530401, National Natural Science Foundation of China Grant Nos. 11404407 and 91121015, Natural Science Foundation of Jiangsu Province No. BK20140072, China Postdoctoral Science Foundation Nos. 2015M580965 and 2016T90028, and the Fondo Nacional de Desarrollo Cient\'ifico y Tecnol\'ogico (FONDECYT, Chile) under grant 1150653.


\renewcommand{\theequation}{A-\arabic{equation}}
\setcounter{equation}{0}  

\renewcommand{\thefigure}{A\arabic{figure}}
\setcounter{figure}{0}  

\begin{widetext}
\section*{Appendix}
\label{sec Appendix}

\subsection{Quantization of the circuit model and its quantum dynamics}

A detailed analysis of circuit quantization of Fig.~\ref{fig5} can be found in Ref.~\cite{felicetti2014}. The full system Hamiltonian that includes the two quantum Rabi models and the resonator-resonator coupling reads
\bal
H=& H_{p,l} + H_{p,r} +\hbar\sum_{j=l,r}\big[\bar{J}_j+J_j(t)\cos(\omega_d t+\varphi_d)\big](a_j+a^{\dag}_j)^2+\hbar\big[\bar{J}_0 + J_0(t)\cos(\omega_d t+\varphi_d)\big](a_l+a^{\dag}_l)(a_r+a^{\dag}_r),
\label{eq_htot5}
\eal
with
\bal
{\bar{J}_j}&=\frac{\phi_0}{4 I_c \cos{\bar\phi}} \frac{\omega_j}{Z^2_j C_j}, & \bar{J}_0&=2\sqrt{\bar{J}_l\bar{J}_r},\nonumber\\
J_j(t) &= \frac{ \phi_0}{4 I_c} \frac{\sin \bar\phi}{\cos^2 \bar\phi} \frac{\omega_j}{Z^2_j C_j}\Delta\phi\Omega(t), & J_0(t)&=2\sqrt{J_l(t)J_r(t)}.\nonumber\\
\eal

The above circuit Hamiltonian is obtained by considering a weak harmonic magnetic flux with frequency $\omega_d$ applied to the SQUID, that is,
\beq
\label{eq_fluxtime}
\Phi_e(t) = \bar\Phi + \Delta\Phi \,\Omega(t) \cos{\para{\omega_d t + \varphi_d}},
\eeq
with $\Delta\Phi \ll \bar\Phi$ and $\Omega(t)$ being the normalized temporal envelope of the applied flux. It implies $\bar\phi = \pi \bar\Phi/ \Phi_0$ and $\Delta\phi = \pi \Delta\Phi/ \Phi_0$.

Notice that the above Hamiltonian~\eq{\ref{eq_htot5}} is different from the Hamiltonian discussed in the main text~(\ref{eq_fullh1}). In particular, except the time-depentdent  resonator-resonator coupling terms, the Hamiltonian~(\ref{eq_htot5}) also includes single-mode squeezing terms and time-independent coupling terms. Nonetheless, one can demonstrate that \eq{\ref{eq_htot5}} reduces to \eq{\ref{eq_2rabiip1}} if we consider realistic system parameters. In order to see this correspondence, let us assume identical resonators, $\omega_l=\omega_r=\omega_c$, with impedances $Z_j=80~\Omega$ and capacitances $C_j=200~\rm fF$. For the critical current of the SQUID we consider $I_c=180~\mu \rm{A}$. Notice that recent experiments in circuit QED have considered critical currents in the range of $1\!-\!2~\mu \rm{A}$~\cite{Chris2011}, however, one can increase the critical current by to a larger value or even a few orders of magnitude into the $m\rm{A}$ regime by making the junctions bigger, this would involve a trilayer fabrication process~\cite{Chris_private,Chris2001}. Moreover, we consider a flat ($\Omega(t)=1$) magnetic flux pulse applied to the SQUID with parameters $\bar{\phi}=\pi/4$ and $\Delta\phi=0.1\bar{\phi}$. The reduced flux quantum is $\phi_0=\hbar/(2e)=3.2911\times10^{-16}~\rm Wb$. These parameters lead to $\bar{J}_j\approx5\times10^{-4}\omega_c$, $\bar{J}_0\approx10^{-3}\omega_c$, $J_j\approx4\times10^{-5}\omega_c$, and $J_0\approx8\times10^{-5}\omega_c$.

\bfig[b]
\centering
\includegraphics[scale=0.5]{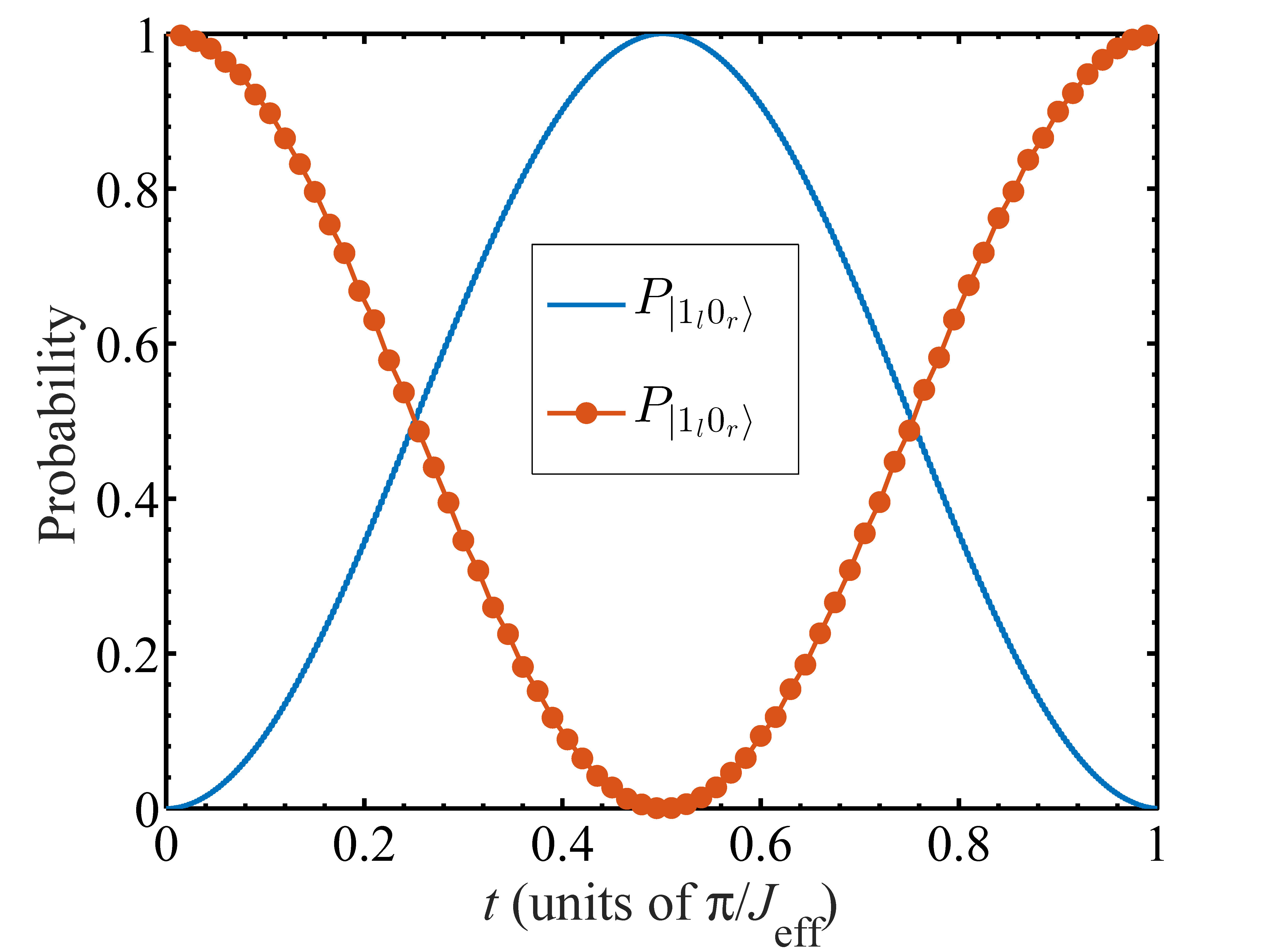}
\caption{Population inversion between states $\ket{1_l0_r}$ and $\ket{0_l1_r}$ for non identical quantum Rabi systems with parameters $\omega_l=1$, $\omega_{a,l}=0.8\omega_l$, $g_l=0.3\omega_l$ for the left quantum Rabi system, and $\omega_r=1$, $\omega_{a,r}=\omega_r$, $g_r=0.9\omega_r$ for the right quantum Rabi system. These values lead to an effective cavity-cavity coupling strength $J_{\rm eff}\approx5.5\times10^{-4}\omega_r$. This simulation has been performed with the full Hamiltonian~(\ref{eq_htot5}) through the Runge-Kutta algorithm.}
\label{Fig7}
\efig

Now we consider the completeness relation for both quantum Rabi systems so that (\ref{eq_htot5}) can be expressed as
\bal
H =& \sum_{{s}=0}  \big(\hbar\omega_{{s},l} \ket{{s}_l}\bra{{s}_l} +\hbar\omega_{{s},r} \ket{{s}_r}\bra{{s}_r} \big) +\sum_{j=l,r}\big[\bar{J}_j+J_j(t)\cos(\omega_dt+\varphi_d)\big]\Bigg\{\sum_{{s}_j,{t}_j>{s}_j}\big(X_{{s}_j{t}_j}\ket{{s}_j}\bra{{t}_j}+X^*_{{s}_j{t}_j}\ket{{t}_j}\bra{{s}_j}\big)+\sum_{{s}_j}X_{{s}_j{s}_j}\ket{{s}_j}\bra{{s}_j}\Bigg\}\nonumber\\
&+\big[\bar{J}_0 + J_0(t)\cos(\omega_dt+\varphi_d)\big]\Bigg\{\sum_{{s}_l,{t}_l>{s}_l}\big(f_{{s}_l{t}_l}\ket{{s}_l}\bra{{t}_l} + {\rm H.c.} \big)\otimes\sum_{{u}_r,{v}_r>{u}_r}\big(f^r_{{u}_r{v}_r}\ket{{u}_r}\bra{{v}_r}  + {\rm H.c.}\big)\Bigg\},
\label{eq_htot6}
\eal
where $X_{{s}_j{t}_j}=\bra{{s}_j}(a_j+a^{\dag}_j)^2\ket{{t}_j}$ and $f_{{s}_j{t}_j}=\bra{{s}_j}(a_j+a^{\dag}_j)\ket{{t}_j}$. Notice that the single mode squeezing operator $(a_j+a^{\dag}_j)^2$ is an even operator according with the parity symmetry of the system. It means that will connect states within the same parity subspace. Now, if we go to an interaction picture with respect to $H_{p,l}+H_{p,r}$, the Hamiltonian~(\ref{eq_htot6}) reads
\bal
H_I(t) =& \sum_{j=l,r}\big[\bar{J}_j+J_j(t)\cos(\omega_dt+\varphi_d)\big]\Bigg\{\sum_{{s}_j,{t}_j>{s}_j}\big(e^{-i\omega_{{t}_j{s}_j} t}X_{{s}_j{t}_j}\ket{{s}_j}\bra{{t}_j}+e^{i\omega_{{t}_j{s}_j} t}X^*_{{s}_j{t}_j}\ket{{t}_j}\bra{{s}_j}\big)+\sum_{{s}_j}X_{{s}_j{s}_j}\ket{{s}_j}\bra{{s}_j}\Bigg\}\nonumber\\
&+\big[\bar{J}_0 + J_0(t)\cos(\omega_dt+\varphi_d)\big]\Bigg\{\sum_{{s}_l,{t}_l>{s}_l}\big(e^{-i\omega_{{t}_l{s}_l} t}f_{{s}_l{t}_l}\ket{{s}_l}\bra{{t}_l} + {\rm H.c.} \big)\otimes\sum_{{u}_r,{v}_r>{u}_r}\big(e^{-i\omega_{{t}_r{s}_r} t}f_{{u}_r{v}_r}\ket{{u}_r}\bra{{v}_r}  + {\rm H.c.}\big)\Bigg\}.
\label{eq_htot7}
\eal
If we restrict the three lowest energy levels for each quantum Rabi system, the above Hamiltonian reads
\bal
H_I(t) =& \sum_{j=l,r}\big[\bar{J}_j+J_j\cos(\omega_dt+\varphi_d)\big]\Bigg\{\big(e^{-i\omega_{{2}_j{1}_j} t}X_{{1}_j{2}_j}\ket{{1}_j}\bra{{2}_j}+e^{i\omega_{{2}_j{1}_j} t}X^*_{{1}_j{2}_j}\ket{{2}_j}\bra{{1}_j}\big)+\sum_{{s}_j=0}^{2}X_{{s}_j{s}_j}\ket{{s}_j}\bra{{s}_j}\Bigg\}\nonumber\\
&+\Big[\bar{J}_0 + J_0\cos(\omega_dt+\varphi_d)\Big]\Bigg(e^{-i\omega_{{1}_l{0}_l} t}f_{{0}_l{1}_l}\ket{{0}_l}\bra{{1}_l}+e^{-i\omega_{{2}_l{0}_l} t}f_{{0}_l{2}_l}\ket{{0}_l}\bra{{2}_l} + {\rm H.c.} \Bigg)\otimes\Bigg(e^{-i\omega_{{1}_r{0}_r} t}f_{{0}_r{1}_r}\ket{{0}_r}\bra{{1}_r}+e^{-i\omega_{{2}_r{0}_r} t}f_{{0}_r{2}_r}\ket{{0}_r}\bra{{2}_r} + {\rm H.c.} \Bigg).
\label{eq_htot7}
\eal

For non identical quantum Rabi systems, the above Hamiltonian can produce single excitations transfer if the resonance condition for the driving frequency is $\omega_d=\omega_{1_r0_r}+\bar{J}_j(X_{1_r1_r}-X_{0_r0_r})-[\omega_{1_l0_l}+\bar{J}_j(X_{1_l1_l}-X_{0_l0_l})]$. Notice that single-mode squeezing terms $\bar{J}_jX_{s_js_j}$ produce energy shifts for each dressed state. Furthermore, the rotating-wave approximation holds under conditions $\bar{J}_j|X_{1_j2_j}|\ll\omega_{2_j1_j}$, $J_j|X_{1_j2_j}|\ll|\omega_{2_j1_j}\pm\omega_d|$, $J_j|X_{s_js_j}|\ll\omega_d$, and $\bar{J}_0 f^{*}_{0_l1_l}f_{0_r1_r}\ll|\omega_{1_l0_l}-\omega_{1_r0_r}|$. The effective coupling strength between the two specific dressed states of the two quantum Rabi models is given by $J_{\rm eff}=(J_0/2)f^{*}_{0_l1_l}f_{0_r1_r}$.

We have performed {\it ab initio} numerics by considering the Hamiltonian~(\ref{eq_htot5}). Figure \ref{Fig7} shows the population inversion between states $\ket{1_l0_r}$ and $\ket{0_l1_r}$ for non identical quantum Rabi systems, see the caption to check parameters. For simplicity we have taken $\varphi_d=0$. It is quite clear the correspondence with the Hamiltonian~(\ref{eq_2rabi3}) discussed in the main text.
\end{widetext}

\end{document}